\documentclass[aps,prb,reprint,superscriptaddress,preprintnumber,showpacs]{revtex4-1}
\usepackage{graphicx}
\usepackage{color}
\usepackage{dcolumn}

\begin{document}


\title{Two dimensional Dirac fermions and quantum magnetoresistance in CaMnBi$_2$}
\author{Kefeng Wang}
\affiliation{Condensed Matter Physics and Materials Science Department, Brookhaven National Laboratory, Upton New York 11973 USA}
\author{D. Graf}
\affiliation{National High Magnetic Field Laboratory, Florida State University, Tallahassee, Florida 32306-4005, USA}
\author{Limin Wang}
\affiliation{Condensed Matter Physics and Materials Science Department, Brookhaven National Laboratory, Upton New York 11973 USA}
\author{Hechang Lei}
\affiliation{Condensed Matter Physics and Materials Science Department, Brookhaven National Laboratory, Upton New York 11973 USA}
\author{S. W. Tozer}
\affiliation{National High Magnetic Field Laboratory, Florida State University, Tallahassee, Florida 32306-4005, USA}
\author{C. Petrovic}
\affiliation{Condensed Matter Physics and Materials Science Department, Brookhaven National Laboratory, Upton New York 11973 USA}

\date{\today}

\begin{abstract}
We report two dimensional Dirac fermions and quantum magnetoresistance in single crystals of CaMnBi$_2$. The non-zero Berry's phase, small cyclotron resonant mass and first-principle band structure suggest the existence of the Dirac fermions in the Bi square nets. The in-plane transverse magnetoresistance exhibits a crossover at a critical field $B^*$ from semiclassical weak-field $B^2$ dependence to the high-field unsaturated linear magnetoresistance ($\sim 120\%$ in 9 T at 2 K) due to the quantum limit of the Dirac fermions. The temperature dependence of $B^*$ satisfies quadratic behavior, which is attributed to the splitting of linear energy dispersion in high field. Our results demonstrate the existence of two dimensional Dirac fermions in CaMnBi$_2$ with Bi square nets.
\end{abstract}


\maketitle


The magnetoresistance (MR) of condensed matter gives information about the characteristics of the Fermi surface and provides promising candidates for magnetic memory or other spintronic devices.\cite{mr1,mr2,mr3} The normal MR in conventional metals is small because semiclassical transport gives quadratic field-dependent MR in the low field range which would saturate in the high field.\cite{mr3} Application of strong magnetic field ($B$) leads to quantization of the orbital motion and results in quantized Landau levels $E_n$ (LLs). In the extreme quantum limit where only the lowest LL dominates, a large linear MR could be expected.\cite{quantumtransport,agte1,agte2,bi,quantummr} However, the required magnetic field for the quantum limit in metals with parabolic bands is usually very large because LLs are equidistant. The exceptions are large linear MR in Ag$_{2-\delta}$Te/Se and Bi film below 6 T.\cite{agte1,agte2,bi} A. A. Abrikosov proposed that the linear MR is intimately connected with linear energy dispersion~\cite{quantumtransport} and recent first principle calculations confirmed the existence of surface states with linear energy-momentum relationship, the so-called Dirac fermions.\cite{fangzhong}

The distance between the lowest and $1^{st}$ LLs of Dirac fermions in magnetic field is very large. The quantum limit where all of the carriers occupy only the lowest LL is easily realized in relatively small fields.\cite{LL1,LL2} Consequently large linear MR could be achieved. Besides Ag$_{2-\delta}$Te/Se, the unsaturated linear MR and other quantum transport phenomena were experimentally observed in other Dirac materials, such as topological insulators (TIs), graphene and some organic conductors.\cite{qt1,qt2,qt3,qt4,graphene} Recently, highly anisotropic Dirac states were observed in SrMnBi$_2$,\cite{srmnbi2} where linear energy dispersion originates from the crossing of two Bi $6p_{x,y}$ bands in the double-sized Bi square net which is a part of (SrBi)$^{+}$ layer.\cite{srmnbi2,srmnbi22}

In this work, we report the quantum oscillation and magnetoresistant behavior in CaMnBi$_2$ single crystals with different layered structure but similar two dimensional (2D) Bi square nets when compared to SrMnBi$_2$. The non-zero Berry's phase, small cyclotron resonant mass and first-principle band structure suggest the existence of Dirac fermions in the Bi square nets. The quasi-2D in-plane transverse magnetoresistance exhibits a crossover at a critical field $B^*$ from semiclassical weak-field $B^2$ dependence to the high-field linear-field dependence. The temperature dependence of $B^*$ satisfies quadratic behavior attributed to the splitting of linear energy dispersion in high field.

Single crystals of CaMnBi$_2$ were grown using a high-temperature self-flux method.\cite{crystal} Stoichiometric mixtures of Ca (99.99$\%$), Mn (99.9$\%$) and excess Bi (99.99$\%$) with ratio Ca:Mn:Bi=1:1:9 were sealed in a quartz tube, heated to 1050 $^{\circ}C$ and cooled to 450 $^{\circ}C$ where the crystals were decanted. The resultant crystals are plate-like and the basal plane of a cleaved crystal is the crystallographic $ab$-plane. Electrical transport measurements up to 9 T were conducted in Quantum Design PPMS-9 with conventional four-wire method. In the in-plane measurements, the current path was in the \textit{ab}-plane, whereas magnetic field was parallel to the \textit{c}-axis except in the rotator experiments. In the out of plane ($c$-axis) resistivity measurements, electric current and magnetic fields were parallel to the $c$-axis. High field MR oscillation up to 35 T were performed at National High Magnetic Field Laboratory in the same configuration to the in-plane MR. {\bf The magnetization measurements were performed in a Quantum Design MPMS in both zero field cooling (ZFC) and field cooling (FC).} Fist principle electronic structure calculation were performed using experimental lattice parameters within the full-potential linearized augmented plane wave (LAPW) method \cite{wien2k1} implemented in WIEN2k package.\cite{wien2k2} The general gradient approximation (GGA) of Perdew \textit{et al}.,\cite{gga} was used for exchange-correlation potential. 

\begin{figure}[tbp]
\includegraphics[scale=0.4]{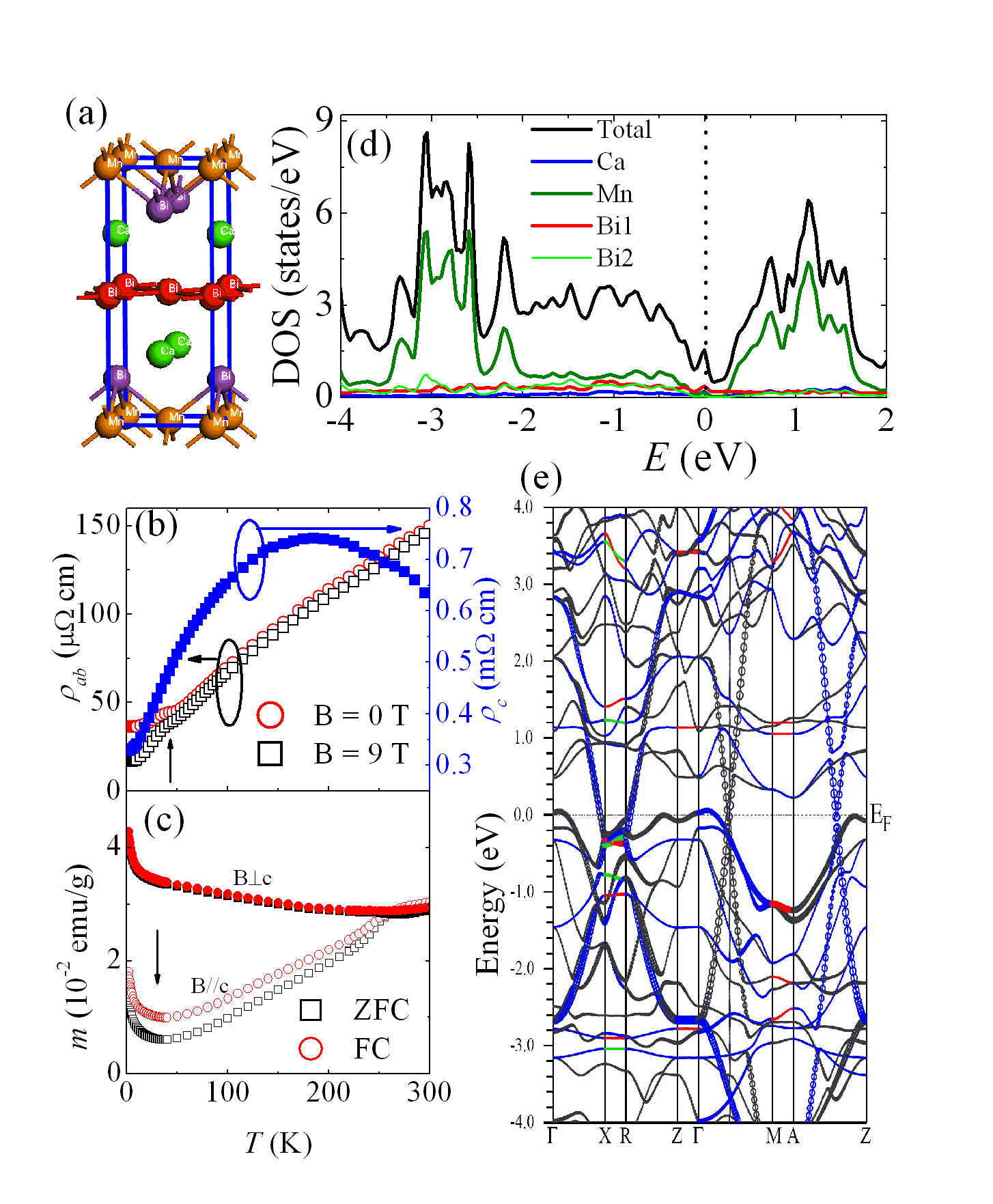}
\caption{(Color online) (a) Crystal structures of CaMnBi$_2$. Bi atoms in  2D square nets are shows by red balls. Ca atoms are denoted by green balls. Another location of Bi atoms is denoted by purple balls. Mn atoms are denoted by orange balls. Blue lines define the unit cell. (b) Temperature dependence of the in-plane resistivity $\protect\rho_{ab}(T)$ (open symbols) and $c$-axis resistivity $\rho_c(T)$ (filled symbols) of the CaMnBi$_2$ single crystal in $B=0$ T (squares) and $B=9$ T (circles) magnetic field respectively. (c) Magnetization ($M$) in $B=1$ T field applied parallel (open symbols) and perpendicular (filled symbols) to the $c$-axis in both ZFC (squares) and field cooling FC (circles) runs. (d) The total DOS (black line) and local DOS from Ca (blue line), Mn (olive line), Bi square nets (Bi1, red line) and Bi in MnBi$_4$ tetrahedron (Bi2, green line) for AFM CaMnBi$_2$. The dotted line indicates the position of the Fermi energy. (e) The band structure for CaMnBi$_2$. The heavy lines with circles denote the bands from Bi square nets and the dotted line indicates the position of the Fermi energy.}
\end{figure}


CaMnBi$_2$ unit cell with P4/nmmm space group contains alternatively stacked two MnBi$_4$ tetrahedron layers and a 2D Bi square net separated by Ca atoms along the $c$-axis (Fig. 1(a)). The MnBi$_4$ tetrahedrons are less distorted and the lattice is smaller when compared to SrMnBi$_2$ with I4/mmm space group since Ca has smaller radius than Sr.\cite{srmnbi2,lattice} The in-plane resistivity $\rho_{ab}(T)$ (Fig. 1(b)) is metallic with a weak anomaly at $\sim 50$ K. The resistivity along the $c$-axis is higher in magnitude than the in-plane resistivity with $\rho_c(T)/\rho_{ab}(T)\sim10-15$ below 100 K. In what follows we will only discuss the in-plane MR. An external magnetic field enhances the low-temperature resistivity and the MR ratio MR$=(\rho_{ab}(B)-\rho_{ab}(0))/\rho_{ab}(0)$ reaches $120\%$ at 2 K in 9 T field. As the temperature is increased, the magnetoresistance is gradually suppressed and becomes negligible above $\sim 50$ K. Magnetization shows a kink at $\sim250$ K indicating an antiferromagnetic (AFM) transition (Fig. 1(d)). The anomaly in the resistivity at ~ 50 K is clearly not related to the AFM order and is possibly due to the weak ferromagnetic order or impurity scattering since the magnetization shows an upturn around that temperature, which is denoted by arrow in Fig. 1(c).

The spin polarized first-principle calculation reveals that the net magnetization in the unit cell is nearly zero confirming the AFM ground state (Fig. 1(d) and (e)). The structure with N$\acute{e}e$l-type AFM configuration in the $ab$-plane has the lowest energy in first-principle calculation because the MnBi$_4$ layers are separated by Ca and Bi layers along the $c$-axis and consequently the inter-layer coupling is rather weak. The density of states (DOS) at the Fermi level of CaMnBi$_2$ (Fig. 1(d)) is dominated by the contribution from states in Bi square nets since the AFM order of Mn ions expels the states of Mn away from the Fermi level. The band structure (Fig. 1(e)) confirms this. There are two narrow bands at the Fermi level with nearly linear energy dispersion along the $\Gamma-M$ and the $A-Z$ directions in addition to a wide band along the $X-R$ direction. Hence the 2D Bi square nets of CaMnBi$_2$ host Dirac states with quasi-2D Fermi surfaces (FS).


\begin{figure}[tbp]
\includegraphics [scale=0.3]{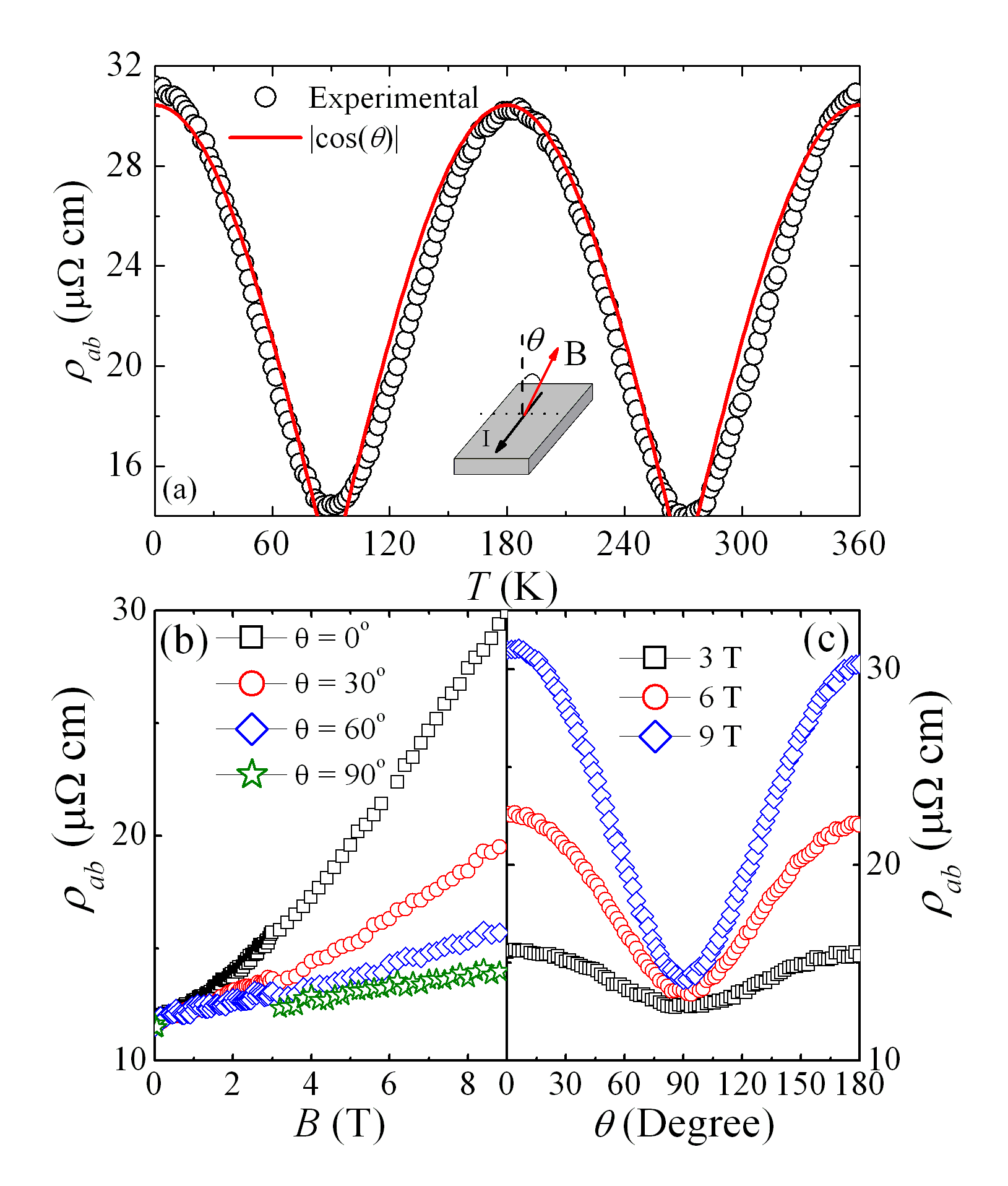}
\caption{(Color online) (a) In-plane resistivity $\rho$ vs. the tilt angle $\protect\theta$ from $0^o$ to $
360^o$ at $B$ = 9 T and $T$ = 2 K for CaMnBi$_2$. The red solid line is the fitting curve using $|\cos(
\protect\theta)|$ (see text). Inset shows the configuration of the measurement. (b) In-plane Resistivity $\protect\rho$ vs. magnetic field $B$ of
CaMnBi$_2$ crystal with different tilt angle $\protect\theta$ between magnetic field and sample
surface ($ab$-plane) at 2 K. (c) $\protect\rho$ vs. the tilt
angle $\protect\theta$ in the fixed magnetic fields (3 T, 6 T and 9 T) and 2
K.}
\end{figure}

The magnetotransport of solids only responds to the extremal cross section $S_F$ of the Fermi surface along the field direction. For a (quasi-) 2D FS, the cross section has $S_F(\theta)=S_0/|\cos(\theta)|$ angular dependence, and 2D states will only respond to perpendicular component of the magnetic field $B|\cos(\theta)|$.\cite{mr3} For example, the 2D states in graphene and the surface states of TIs exhibit $|\cos\theta|$ angular dependent magnetotransport.\cite{LL2,qt1} The magnetoresistance of CaMnBi$_2$ shows significant dependence on the field direction (Fig. 2). The crystal was mounted on a rotating stage such that the tilt angle $\theta$ between the crystal surface ($ab$-plane) and the magnetic field can be continuously changed with currents flowing in the $ab$-plane perpendicular to magnetic field, as shown in the inset of Fig. 2(a).  Angular dependent magnetoresistance $\rho(B,\theta)$ at $T\sim 2$ K is shown in Fig. 2(b) and (c). When $B$ is parallel to the $c$-axis ($\theta=0^{o}, 180^{o}$), the MR is maximized and is linear in field for high fields. With increase in the tilt angle $\theta$, the MR gradually decreases and becomes nearly negligible for
$B$ in the $ab$-plane ($\theta=90^o$). Angular dependent resistivity in $B=9$ T and $T=2$ K shows wide maximum when the field is parallel to the $c$-axis ($\theta=0^o, 180^o$), and sharper minimum around $\theta=90^o, 270^o$ (Fig. 2(a)). The whole curve follows the function of $|\cos(\theta)|$ very well (red line in Fig. 2(a)). The angular dependent in-plane magnetoresistance suggests the quasi-2D Fermi surface.\cite{mr3}

\begin{figure}[tbp]
\includegraphics[scale=0.3]{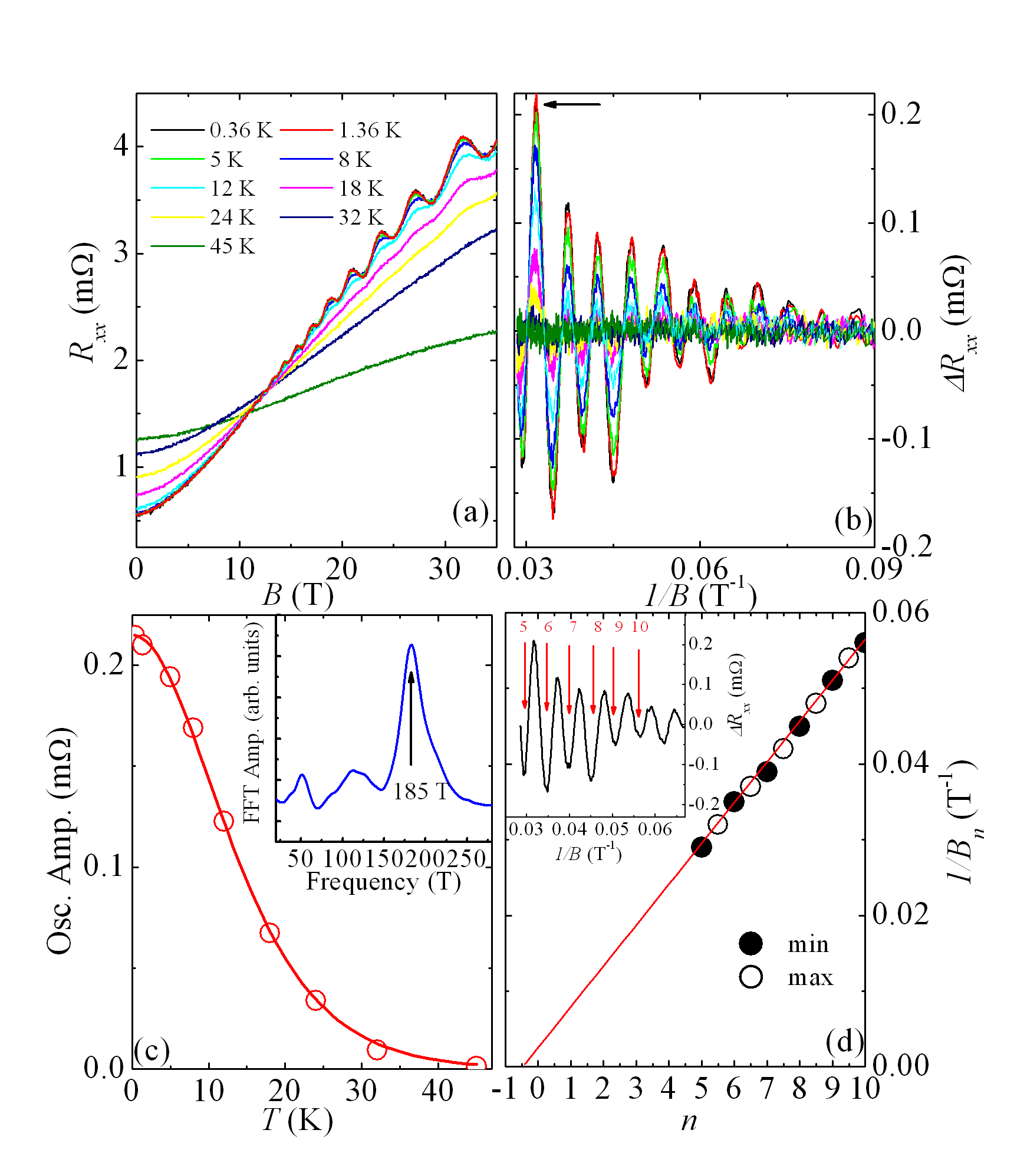}
\caption{(Color online) (a) Magnetic field dependence of resistance of CaMnBi$_2$ crystal and (b) Magnetoresistant SdH oscillations $\Delta R_{xx}=R_{xx}-<R_{xx}>$ as a function of $1/B$ below 35 T. (c) Temperature dependence of the oscillation amplitude (Osc. Amp.) in magnetoresistant SdH oscillations. The red line is the fitting results giving cyclotron mass. Inset shows the Fourier transform spectrum of the SdH oscillation. (d) The integer Landau levels as a function of inverse field. Inset shows the oscillation of $\Delta R_{xx}$ at 0.36 K where arrows indicate the positions of estimated LL index $n$ labeled by the numbers.}
\end{figure}

\begin{figure}[tbp]
\includegraphics [scale=0.4]{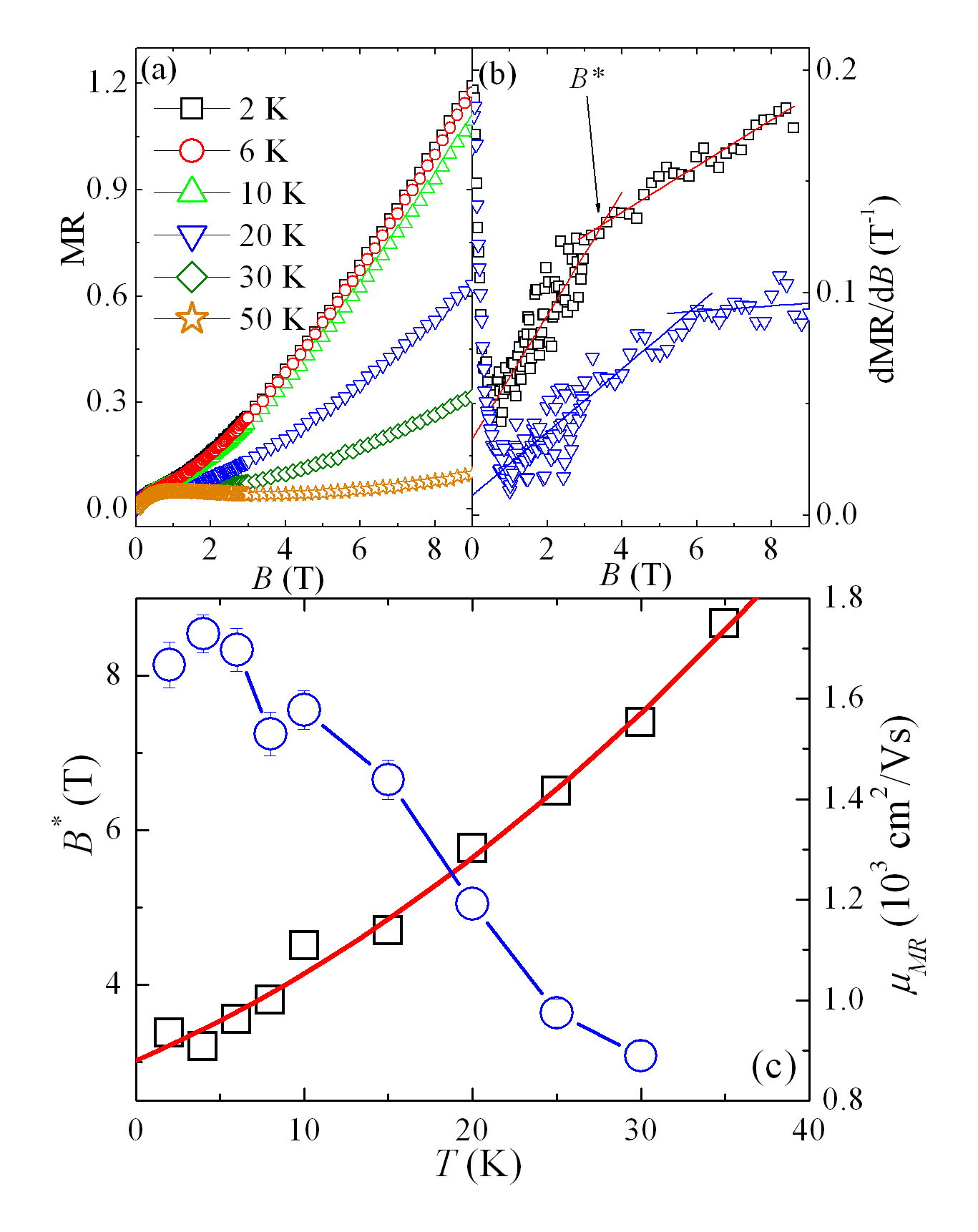}
\caption{(Color online) (a) The magnetic field ($B$) dependence of the in-plane magnetoresistance MR at different
temperatures. (b) The field derivative of in-plane MR at different temperature
respectively. The lines in high field regions were fitting results using MR $=A_1B+O(B^2)$ and the lines in low field regions using MR $=A_2B^2$. (c) Temperature dependence of the critical field $B^*$ (black
squares) and the effective MR mobility $\mu_{MR}$ extracted from the weak-field MR (blue circles). The red solid line is the fitting results of $B^*$ using $B^*=\frac{1}{2e\hbar
v_F^2}(E_F+k_BT)^2$. }
\end{figure}

In Fig. 3(a) and (b), the in-plane magnetoresistance $R_{xx}$ and the $\Delta R_{xx}=R_{xx}-<R_{xx}>$ shows clear Shubnikov-de Hass (SdH) oscillations below 45 K. The Fourier transform spectrum of the oscillation at 0.36 K (inset in Fig. 3(c)) reveals a  periodic behavior in $1/B$ with a frequency $F=185$ T. The temperature dependence of the oscillation amplitude can be used to determine cyclotron effective mass through the Lifshitz-Kosevitch formula.\cite{oscillation} Using the highest oscillation peak (indicated by the arrow in Fig. 3(b)), the fitting gives a $m\approx 0.35 m_e$ where $m_e$ is the bare electron mass (Fig. 3(c)). In metals, SdH oscillations correspond to successive emptying of LLs as the magnetic field is increased. The LL index $n$ is related to the cross section of FS $S_F$ by $2\pi(n+\gamma)=S_F\frac{\hbar}{eB}$. In the index plot (Fig. 3(d)), the inverse peak and minimum fields $1/B$ fall on a straight line (red line) versus the integers $n$ and the extrapolation of the high-field SdH peaks and minimum gives $\gamma\simeq 0.45$. $\gamma$ should be zero for conventional metals but $(\pm)1/2$ for Dirac fermions due to the nonzero Berry's phase associated with their cyclotron motion. The Berry's phase and the $\sim 1/2$ intercept of the linear fit of LLs have been observed in Dirac fermion systems, such as monolayer graphene \cite{LL1} and topological insulators.\cite{qt1,qt2} The resulted $\gamma \sim 1/2$, as well as the small cyclotron mass $0.35m_e$ reveals the presence of Dirac fermions in CaMnBi$_2$.

Fig. 4(a) shows the magnetic field dependence of MR at different temperatures and Fig. 4(b) shows the field derivative of MR, $d$MR$/dB$. $d$MR$/dB$ initially decreases with increase in field indicating $B^{1/2}$ dependence of MR, and then linearly increases with field in the low field region which indicates a $B^2$ dependent MR by linear fitting (lines in the low field region).  But above a characteristic field $B^*$, $d$MR$/dB$ saturates to a much reduced slope  This indicates that in the high fields the MR is dominated by a linear field dependence plus a very small quadratic term (MR $=A_1B+O(B^2)$) as shown by lines in the high-field region. With increase in temperature, MR decreases and the cross over field $B^*$ increases gradually. Above 50 K MR becomes negligible. Below 9 T and 50 K, the evolution of $B^*$ with temperature is parabolic (squares in Fig. 4(c)).

The energy splitting between the lowest and $1^{st}$ LLs of Dirac fermions can described by $\triangle_{LL}=\pm v_F\sqrt{2e\hbar B}$ where $v_F$ is the Fermi velocity.\cite{LL1,LL2,qt1,qt2} In the quantum limit at specific temperature and field, $\triangle_{LL}$ becomes larger than both the Fermi energy $E_F$ and the thermal fluctuations $k_BT$ at a finite temperature. Consequently all carriers occupy the lowest Landau level and eventually the quantum transport with linear magnetoresistance shows up. The critical field $B^*$ above which the quantum limit is satisfied at specific temperature $T$ is $B^*=\frac{1}{2e\hbar v_F^2} (E_F+k_BT)^2$.\cite{qt3} The temperature dependence of critical field $B^*$ in CaMnBi$_2$ clearly deviates from the linear relationship and can be well fitted by the above equation, as shown in Fig. 4(c). This confirms the existence of Dirac fermion states in CaMnBi$_2$.

In a multiband system with both Dirac and conventional parabolic-band carriers (including electrons and holes) where the Dirac carriers are dominant in transport, the coefficient of the low-field semiclassical $B^2$ quadratic term, $A_2$, is related to the effective MR mobility $\sqrt{A_2}=\frac{\sqrt{\sigma_e\sigma_h}}{\sigma_e+\sigma_h}(\mu_e+\mu_h)=\mu_{MR}$ (where $\sigma_e, \sigma_h, \mu_e, \mu_h$ are the effective electron and hole conductivity and mobility in zero field respectively). The effective MR is smaller than the average mobility of carriers $\mu_{ave}=\frac{\mu_e+\mu_h}{2}$ and gives an estimate of the lower bound.\cite{qt3,qt4} Fig. 4(c) shows the dependence of $\mu_{MR}$ on the temperature. At 2 K, the value of $\mu_{MR}$ is about 1800 cm$^2$/Vs in CaMnBi$_2$ which is larger than the values in conventional metals.

Compared to SrMnBi$_2$, the effective MR mobility in CaMnBi$_2$ is smaller ($\sim 3400$ cm$^2$/Vs in SrMnBi$_2$), while the crossover field $B^*\sim 3$ T at 2 K and the cyclotron mass $m\sim 0.35m_e$ is larger, implying smaller Fermi velocity of Dirac Fermions. This may be due to the contribution of the wide parabolic band in CaMnBi$_2$ (as shown in Fig. 1(d)) which is absent in SrMnBi$_2$.\cite{srmnbi2,srmnbi22,kefeng} {\bf The Berry's phase revealed by the quantum oscillation, combined with our first-principle electronic structure and the quadratic-temperature dependence of the crossover field from semi-classical transport to quantum linear magnetoresistance is more than convincing evidence for the existence of Dirac fermions in 2D Bi suqare nets of CaMnBi$_2$ and SrMnBi$_2$.\cite{kefeng} The direct obervation of the linear energy dispersion in Bi square nets and the detailed information on multiband characteristics could deserve further study by more powerful spectroscopy methods such as angle-resolved photoemission spectroscopy (ARPES) and will be sought after.}


In summary, we report two dimensional Dirac fermions and quantum magnetoresistance in single crystals of CaMnBi$_2$. The non-zero Berry's phase, small cyclotron resonant mass and first-principle band structure suggest the existence of 2D Dirac fermions in the Bi square nets. The in-plane transverse magnetoresistance exhibits a crossover at a critical field $B^*$ from semiclassical weak-field $B^2$ dependence to the high-field unsaturated linear magnetoresistance ($\sim 120\%$ in 9 T at 2 K) due to the quantum limit of the 2D Dirac fermions. The temperature dependence of $B^*$ satisfies quadratic behavior, which is attributed to the splitting of linear energy dispersion in high field. Our results demonstrate the existence of two dimensional Dirac fermions in CaMnBi$_2$ with similar Bi square net structural unites.

\begin{acknowledgments}
We thank John Warren for help with SEM measurements. This work was performed at Brookhaven and supported by the U.S. DOE under contract No. DE-AC02-98CH10886. Work at the National High Magnetic Field Laboratory is supported by the DOE NNSA DE-FG52-10NA29659 (S. W. T and D. G.), by the NSF Cooperative Agreement No. DMR-0654118 and by the state of Florida.
\end{acknowledgments}


\end{document}